\documentclass[showpacs,twocolumn,aps,prl,superscriptaddress]{revtex4}

\usepackage{mathrsfs}
\usepackage{amsmath}
\usepackage{amsfonts}
\usepackage{amssymb}
\usepackage{amsthm}
\usepackage{graphicx}

\begin{document}
\title{New Galvanomagnetic Effects of Polycrystalline
Magnetic Films}

\author{Y. Zhang}
\affiliation{Physics Department, The Hong Kong University of
Science and Technology, Clear Water Bay, Kowloon, Hong Kong}
\affiliation{HKUST Shenzhen Research Institute, Shenzhen 518057, China}
\author{H. W. Zhang}
\affiliation{School of Microelectronics and Solid-State Electronics,
University of Electronic Science and Technology of China, Chengdu,
Sichuan 610054, China}
\author{X. R. Wang}
\email{phxwan@ust.hk}
\affiliation{Physics Department, The Hong Kong University of
Science and Technology, Clear Water Bay, Kowloon, Hong Kong}
\affiliation{HKUST Shenzhen Research Institute, Shenzhen 518057, China}

\begin{abstract}
Within the linear response of polycrystalline magnetic films to electric
currents, a general analysis predicts three new galvanomagnetic
effects originated from the two-dimensional nature of the films.
These new galvanomagnetic effects, which differ from the conventional
extraordinary Hall effect and anisotropic magnetoresistance, are follows.
1) The longitudinal resistivity depends linearly on the magnetization
component perpendicular to a film. 2) A current parallel to the
magnetization can generate an electric field in the vertical direction
of a film. 3) A current perpendicular to a film can generate an
electric field along the magnetization direction.
\end{abstract}

\maketitle



The galvanomagnetic effects are a well-studied subject in
condensed matter physics \cite{MacDonald,conwell,ganichev}.
The subject is about the generation of electric field in a
metal or a semiconductors by an electric current.
These effects are often used to probe the electronic properties of matter.
The galvanomagnetic effects include many well-known phenomena such as
classical and quantum Hall effect \cite{QHE}, extraordinary Hall effect
(EHE) \cite{pugh,luttinger,chazalviel,Niu}, anisotropic magnetoresistance
(AMR) \cite{AMR,Tang}, and de Haas-Schubnikov oscillations \cite{dhs},
as well as recently discovered spin Hall and inverse spin Hall
effect \cite{hirsch, hoffmann, kimura, kajiwara, saitoh}.
Within the linear response region, the theories of galvanomagnetic effects
are well-developed. For example, one can use the Kubo formula \cite{kubo}
or Boltzmann theory to compute resistivity tensor of a given microscopic
model in the presence or absence of a magnetic field. One can also use
semiclassical dynamical equations for quasi-electrons or holes
to obtain a current density under given external driven forces.
Although these theories can be used to compute the transport properties
of a given microscopic model, the microscopic mechanism of one particular
galvanomagnetic effect may not be as transparent as one might think.
For example, the origins of the EHE were completely known only in recent
years \cite{Niu}. A microscopic understanding of AMR is still a research
subject. So far, most of the well-known galvanomagnetic effects were
first found in experiments instead of being predicted by theories.
In this letter, new galvanomagnetic effects for polycrystalline magnetic
films are predicted. Our predictions are based on the general requirement
that all physical quantities must be tensors such as scalars (tensors of
rank 0) and vectors (tensors of rank 1), and all physics equations can be
written in tensor forms. Our analysis shows that all possible galvanomagnetic
effects for polycrystalline magnetic bulks are already known.
However, three new galvanomagnetic effects, which are not known to the
best of our knowledge, exist in polycrystalline magnetic films due to their
two dimensional nature.

Consider a piece of ferromagnetic metal with a current density
$\mathbf J$ passing through it.
The galvanomagnetic effects of the system in the linear response
region is best described by the generalized Ohm's law.
The well-known generalized Ohm's law of polycrystalline ferromagnetic
metals is \cite{juretschke, egan, gui}
\begin{equation}
\mathbf E=\rho_\perp\mathbf J+\frac{\Delta\rho}{M^2}(\mathbf J\cdot\mathbf
M)\mathbf M-R_0\mathbf J\times\mathbf H-R_1\mathbf J\times\mathbf M,
\label{ohms_law}
\end{equation}
where $M$ is the magnitude of magnetization $\mathbf M$ and $\mathbf H$
is an applied external magnetic field. $\mathbf E$ is the electric field
induced by the current density $\mathbf J$.
$\rho_{\perp}$ is the resistivity when $\mathbf M$ and $\mathbf J$ are
perpendicular with each other. $\Delta\rho=\rho_{||}-\rho_{\perp}$ is
the difference between $\rho_{\perp}$ and the resistivity $\rho_{||}$ when
$\mathbf M$ is parallel to $\mathbf J$. This term is usually called the AMR.
This AMR gives the famous universal angular dependence of longitudinal
resistance (resistivity) $\rho_{xx}=\rho_{\perp}+\Delta\rho\cos^2\theta$,
where $\theta$ is the angle between $\mathbf M$ and $\mathbf J$.
$R_0$ and $R_1$ are the ordinary and extraordinary Hall coefficients.
This generalized Ohm's law is the basis of the electrical detection of
ferromagnetic resonance \cite{juretschke, egan, saitoh, gui}.

Eq. \eqref{ohms_law} is in fact the most general linear response of a
polycrystalline magnet. It could be understood by the following reasoning.
For simplicity, let us assume that there is no external magnetic field and
$\mathbf M$ is the only vector variable available in the system \cite{note1}.
For the linear response of the system to a current density $\mathbf J$,
the most general expression of the induced electric field must be
$\mathbf E=\tensor\rho(\mathbf M)\mathbf J$, where $\tensor\rho(\mathbf M)$
is a rank-2 Cartesian tensor, depending on $\mathbf M$.
It is well-known that a Cartesian tensor of rank 2 can be decomposed
into the direct sum of a scalar of function of $M$, a vector that is a
function of $M$ multiplying $\mathbf M$, and a traceless symmetric
tensor that is a function of $M$ multiplying $\mathbf M\mathbf M-M^2/3$.
Thus the most general expression of $\mathbf E$ is
\begin{equation}
\mathbf E=\big (\rho_{\perp}+\frac{\Delta\rho}{3}\big )\mathbf J+R_1
\mathbf M\times\mathbf J +\frac{\Delta\rho}{M^2}\big (\mathbf M \mathbf M
- \frac{M^2}{3}\big )\cdot\mathbf J.
\label{expansion1}
\end{equation}
This is exactly Eq. \eqref{ohms_law}. Although no new physics is
obtained from this reasoning, this analysis is capable of finding all
galvanomagnetic effects for the bulk of polycrystalline magnets.
Interestingly, a similar spirit, but with a much more tedious and
lengthy argument, was used before to obtain the AMR \cite{Ranieri}.

Encouraged by the above success, we carry out a similar analysis for
polycrystalline magnetic films, lying in the $xy$-plane as shown in Fig. 1.
Although a polycrystalline film is isotropic in the film plane,
$\hat z$ is an available vector and $\tensor\rho(\mathbf M,\hat z)$
should be a function of both $\mathbf M$ and $\hat z$.
Of course, the sense of the positive $z$-direction must be determined by the
current density and the in-plane component of the magnetization \cite{note2}.
Since three vectors ($\mathbf M$, $\hat z$, and $\mathbf M\times\hat z$)
and three traceless symmetric tensors ($\mathbf M \mathbf M-M^2/3$,
$\mathbf M \hat z+\hat z \mathbf M-2M_z/3$, and $\hat z\hat z-1/3$) can
be constructed out of $\mathbf M$ and $\hat z$, we have, with a similar
reasoning as that for a bulk polycrystalline magnet,
\begin{equation}
\begin{aligned}
&\mathbf E=\big (\rho_\perp +\frac{\Delta\rho+\rho_2}{3}\big ) \mathbf J+
\big (R_1 \mathbf M+ \\
&\rho_1\hat z+ R_2' \mathbf M\times\hat z\big )\times\mathbf J+ \bigg [
\frac{\Delta\rho}{M^2}\big (\mathbf M\mathbf M-\frac{M^2}{3}\big )+ \\
& R_3'\big (\mathbf M\hat z +\hat z\mathbf M -2\frac{M_z}{3}\big ) +
\rho_2 \big (\hat z\hat z-\frac{1}{3}\big )\bigg ] \cdot\mathbf J= \\
&\rho_\perp \mathbf J-R_1 \mathbf J \times \mathbf M+
\frac{\Delta\rho}{M^2} \big (\mathbf M\cdot\mathbf J\big )\mathbf M-
\rho_1\mathbf J \times \hat z +\\
&R_2J_z\mathbf M+R_3\big (\mathbf M\cdot\mathbf J\big )\hat z -R_4
M_z \mathbf J + \rho_2 J_z\hat z ,
\end{aligned}
\label{expansion2}
\end{equation}
where
$R_2\equiv R_2'+R_3'$, $R_3\equiv R_3'-R_2'$ and $R_4\equiv (2/3)R_2'$.
The $\rho_1$-term may be interpreted as the spin-Hall term if the sign of
$\rho_1$ for spin-up electrons is opposite to that for spin-down electrons.
Microscopically, it is known that spin-orbit interaction can lead to a term
like this \cite{hirsch, hoffmann}. Obviously, $\rho_2$ is the resistivity
along the $z-$direction. Interestingly, one obtains three new terms.
$R_2$-term says that a current perpendicular to the film induces an electric
field in $\mathbf M$ direction. $R_3$-term says that a current along $\mathbf
M$ direction generates an electric field in the $z-$direction. $R_4$-term
says that the longitudinal resistivity of the film depends on $M_z$ linearly.
Since the sense of the positive $z-$direction is lost when the in-plane
components of $\mathbf M$ and $\mathbf J$ are parallel to each other, one
should expect $R_4=0$ when $\mathbf J$ is in-plane. In general, $R_4\neq 0$,
and the longitudinal resistivity of a polycrystalline film changes when
the current reverses its direction so $+\hat z$ becomes $-\hat z$.
This is the fingerprint feature of this term. It destroys the symmetry of
$\mathbf E(\mathbf M, -\mathbf J)=-\mathbf E(\mathbf M, \mathbf J)$ that
Eq. \eqref{ohms_law} posses. Like a semiconductor diode, this term leads to
rectification effect. It should be pointed out that all the coefficients of
these terms depend, in principle, on $M$ as well as other possible variables
or parameters like the temperature and impurities (that will not change our
tensor analysis). Generally speaking, all these new terms should exist.
Of course, their values depend on microscopic interactions that lead to
these terms.

\begin{figure}
\centering
\includegraphics[width=0.45\textwidth]{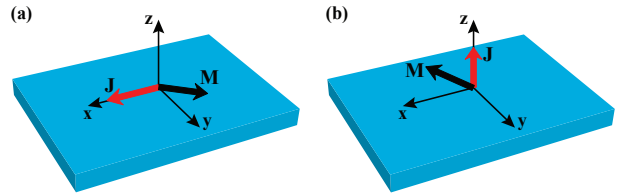}\\
\caption{(Color online) Schematic illustration of two possible experimental
configurations of a polycrystalline magnetic film lying in the $xy-$plane.
The $z$-axis is perpendicular to the film.
(a) Current density $\mathbf J$ flows along the $x$-direction in
the magnetic film. The magnetization $\mathbf M$ is in the $yz-$plane.
(b) Current density $\mathbf J$ flows along the $z$-direction.
The magnetization $\mathbf M$ is in the $xz-$plane.}
\label{configs}
\end{figure}

In order to find out how to experimentally verify the new effects, let
us examine two experimental configurations involving a polycrystalline
magnetic film as schematically shown in Fig. \ref{configs}.
Fig. \ref{configs}(a) is the case that a current density
flows in the film, say $\mathbf J =J \hat x$, while Fig. \ref{configs}(b)
corresponds to a current perpendicular to the film, $\mathbf J =J \hat z$.
In the case of Fig. \ref{configs}(a), electric field $\mathbf E$ in the $x-$,
$y-$ and $z-$directions are proportional to $\rho_{xx}$, $\rho_{yx}$ and
$\rho_{zx}$. According to Eq. \eqref{expansion2}, $\rho_{xx}=\rho_\perp+
(\Delta\rho/M^2)M_x^2-R_4M_z$. The first two terms describe the AMR.
The third term says that the longitudinal resistance has an additional
contribution linear in $M_z$. This contribution vanishes when $\mathbf M$
lies in-plane of the magnetic film. All terms in $\rho_{yx}$, $\rho_{yx}=
(\Delta\rho/M^2)M_xM_y+\rho_1 +R_1M_z$, are known before (AMR, EHE and SHE).
However, $\rho_{zx}=(\Delta\rho/M^2)M_xM_z-R_1M_y+R_3M_x$ depends on $R_3$.
For an in-plane magnetized film, $\rho_{zx}=-R_1M_y+R_3M_x$, so that the
fingerprint of the new terms is $R_3M \cos (\theta)$ behavior, where
$\theta$ is the angle between $\mathbf M$ and $\mathbf J$. Unfortunately, the
film thickness cannot be too large since Eq. (3) is for a two-dimensional
film. A reasonable length scale is electron mean-free path that
is order of nanometers to microns for most metals at normal conditions.
On the other hand, the way to measure a resistivity is to convert it to a
voltage signal. From this consideration, $\rho_{zx}$ is not a good quantity
to measure because one expects voltage signal in vertical direction induced
by an in-plane current would be too small to measure for a micron thick film.
Nevertheless, one can use this configuration to measure $R_4$ by using a
single domain thin film so that the voltage signal would not be averaged out.
To simplify the analysis, it is better to keep $\mathbf M$ in the $yz$-plane
as illustrate in Fig. \ref{configs}(a). In this configuration,
$\rho_{xx}=\rho_\perp-R_4M_z$. Reverse the current direction, one has
$\rho_{xx}=\rho_\perp+R_4M_z$. Thus the difference in voltage drop for the
two opposite current directions is proportional to $R_4$. One can also study
$M_z$-dependence of the voltage drop in the $x-$direction at a fixed
$\mathbf J$. One may use an external magnetic field to tune $M_z$.

In the case of Fig. \ref{configs}(b) and if one chooses $x-$axis parallel
to the in-plane component of $\mathbf M$ ($M_y=0$), then the $x-$,
$y-$ and $z-$components of $\mathbf E$ are proportional to $\rho_{xz}$,
$\rho_{yz}$ and $\rho_{zz}$. This may be detected by measuring the
voltage drops in the $x-$, $y-$, and $z-$directions when a tunneling
current is passing through a multilayer sample in which a thin single
domain magnetic layer is sandwiched between two non-magnetic layers.
According to Eq. \eqref{expansion2}, $\rho_{xz}=(\Delta\rho/M^2)
M_xM_z+R_2M_x$, $\rho_{yz}=-R_1M_x$ and
$\rho_{zz}=(\rho_\perp+\rho_2)+(\Delta\rho/M^2)M_z^2+(R_2+R_3-R_4)M_z$.
Obviously, the voltage drop in the $x-$direction reverses its sign
when $\mathbf M=M\hat x$ changes to $\mathbf M=-M\hat x$.
The difference of the two voltage drops is proportional to $R_2$.
This is the signature of $R_2$. To obtained $R_3$, one can measure the
tunneling resistances for $\mathbf M=\pm M\hat z$. The difference of
the two tunneling resistances is proportional to $R_2+R_3-R_4$.

The new galvanomagnetic effects reported here come from the two
dimensional nature of films. The interfacial effects are not new in
physics \cite{sakurai,lujie,tsujikawa} and exist in both magnetic and
non-magnetic systems. Our general analysis points out that the new
galvanomagnetic effects should exist in a polycrystalline magnetic film
although it reveals neither microscopic origins of these effects nor
their possible magnitudes. To utilize these new effects as material
probes or to search for materials with larger effects, one needs to
understand how microscopic interactions generate these new effects.
Another interesting problem is to generalize the analysis to
crystalline materials because it can help us to understand those
electron transport measurements on magnetic single crystals.
Since a crystal can provide intrinsic crystal vectors or tensors,
$\tensor\rho(\mathbf M,\hat{x}_1, \hat{x}_2,...)$ shall depend
not only on $\mathbf M$, but also on other tensors or vectors
$(\hat{x}_1, \hat{x}_2,...)$ that a crystal possess.
Then, one needs to construct all possible vectors and traceless
symmetric tensors out of these available variables in
order to find out all possible galvanomagnetic effects.

Clearly, all three new terms involve the coupling between the
magnetization and charge current so that they can generate dc-voltage
in the electrical detection of FMR just like what AMR and EHE do
\cite{juretschke,egan,saitoh,gui}. The electrical detection of FMR
has been used in recent years to extract spin pumping and the spin
Hall angle that measures the strength of both spin Hall effect and
the inverse spin Hall effect \cite{juretschke,egan,saitoh,gui}.
Obviously, the new galvanomagnetic effect should have important
implications on reliability of extracting material parameters like
the spin Hall angle of various metals since most experiments used
polycrystalline magnetic films.

In conclusion, we present a general analysis of the linear response of
polycrystalline magnetic films and predict three new galvanomagnetic effects.
The longitudinal resistivity contains, in general, a term linear in the
magnetization component perpendicular to a film. A current along the
magnetization can induce an electric field perpendicular to a film.
Furthermore, a current perpendicular to a film can induce an electric
field along the magnetization direction. Their implications are discussed.
The typical features of these effects are analyzed and possible ways to
experimentally verify them are suggested.

This work was supported by the National Natural Science Foundation of
China (Grant No. 11374249) as well as Hong Kong RGC Grants No.
163011151 and No. 605413.


\begin{thebibliography}{99}
\bibitem {note1}The spin vector of the itinerary electrons is ignored here.
It could involve in the transport through the spin-orbit interactions.
\bibitem {note2}The $z$-direction should be determined by the current flow
direction and the in-plane component of the magnetization if the current
flows in the plane. One would not be able to distinguish the +$z-$direction
from the -$z-$direction if the in-plane component of $\mathbf M$ is parallel
to current density in this case. For current perpendicular to the film,
the +$z-$direction can be determined by the current flow direction.
\end{thebibliography}
\end{document}